\documentclass[a4paper, twocolumn]{revtex4}
\usepackage{amssymb}
\usepackage{amsmath}
\usepackage{graphicx}
\usepackage[english]{babel}
\usepackage[T1]{fontenc}
\usepackage[utf8]{inputenc}

\graphicspath{{./img/}}

\begin{document}

\newcommand{\Ec}{\mathcal{E}}
\newcommand{\Hc}{\mathcal{H}}
\newcommand{\nn}{\textbf{n}}
\newcommand{\kk}{\textbf{k}}

\title{Ionization in a laser assisted ion-ion collision}
\author{O.~Novak}
\author{R.~Kholodov}
\affiliation{Institute of applied physics of National academy of sciences of Ukraine, Petropavlivska street, 58, 40000 Sumy, Ukraine}

\author{A.~N.~Artemyev}
\affiliation{Institut f\"ur Physik and CINSaT, Universit\"at Kassel, Heinrich-Plett-Stra{\ss}e 40, 34132 Kassel, Germany}

\author{A.~Surzhykov}
\affiliation{Physikalisch-Technische Bundesanstalt, Bundesallee 100, 38116 Braunschweig, Germany
and Technische Universitat Braunschweig, Institut fur Mathematische Physik, Mendelssohnstra{\ss}e 3, D-38106 Braunschweig, Germany}

\author{Th.~St\"ohlker}
\affiliation{Helmholtz Institute Jena, Fr\"obelstieg 3, 07743 Jena, Germany}
\affiliation{GSI Helmholtzzentrum f\"ur Schwerionenforschung, Planckstra{\ss}e 1, 64291 Darmstadt, Germany}
\affiliation{Institute for Optics and Quantum Electronics, Friedrich Schiller University, Max-Wien-Platz 1, 07743 Jena, Germany}

\date{2022}

\begin{abstract}
The ionization of a hydrogen-like heavy ion by impact of a charged projectile under simultaneous irradiation by a short laser pulse is investigated within the non-perturbative approach, based on numerical solutions of the time-dependent Dirac equation. 
Special emphasis is placed on the question of whether the laser- and impact-ionization channels interfere with each other, and how this intereference affects the ionization probability. 
To answer this question we performed detailed calculations for the laser-assisted collisions between hydrogen-like Pb$^{81+}$ and $\alpha$ particles. 
The results of the calculations clearly indicate that for the experimentally relevant set of (collision and laser) parameters, the interference contribution can reach 10 \% and can be easily controlled by varying the laser frequency.
\end{abstract}

\maketitle

\section{Introduction}
\label{sec:intro}
Recent advances in ion accelerator and coherent light facilities open new possibilities for the exploration of the strong field domain.
Indeed, heavy highly-charged ions and intense laser pulses provide a laboratory testbed for investigations of the physics of critical (or even over-critical) electromagnetic fields.
For example, laser facilities being constructed in the frame of Extreme Light Infrastructure (ELI) project~\cite{ELI} promise to reach electric field strength comparable with the Schwinger limit~\cite{QEDSF}.
Increase in maximum pulse power is closely related with shortening of pulse duration, and presently short pulses of only few optical cycles can be generated~\cite{Brabec00, Baltuska03, Lopez05}. 
Another source of strong, even though microscopic, electromagnetic fields are highly charged heavy ions. 
Experiments with heavy ions in a controlled charge state up to bare uranium nuclei are currently performed and planned at the the GSI and  the FAIR (Facility for Antiproton and Ion Research) facility in Darmstadt~\cite{FAIR, Gumberidze05, Stoehlker05, Gumberidze09, Kuehl14}. 
The experiments with merging ion beams counter-propagating to a petawatt laser pulse are also anticipated at these facilities.
The studies of the interaction of fast moving highly--charged ions with laser pulses are also intended at the Gamma Factory in CERN~\cite{Budker20}.
In the Gamma Factory setup, the laser in the rest frame of ions experiences Doppler 
boost of both the field strength and the photon energy to the X-ray range. 

The experiments on laser-ion interactions, that are planned at the GSI/FAIR and Gamma Factory facilities, will provide many novel oportunities for probing both structure and dynamics of highly-charged ions. 
For the latter, of special interest is the study of \textit{laser-assisted} fundamental atomic processes. 
Indeed, while in the non-relativistic (low-energy and low-$Z$) regime the ionization, excitation, capture and charge transfer processes in laser-assisted atomic collisions have been explored in detail, see Refs.~\cite{Madsen02,Hansen03,Voitkiv06,Ciappina06,Ciappina08,Kirchner07,Dominguez14,Dominguez15}, not much is known about relativistic collisions of highly-charged heavy ions. 
The storage-ring experiments on ion-ion (or ion-atom) collisions in the presence of intense laser radiation may provide, therefore, more insight into the electron dynamics in the strong field regime. 

This paper is focused on the ionization, that is one of the most fundamental processes in atomic physics. 
The ionization of heavy highly-charged ions is mediated either by relativistic collisions%
~\cite{Wille86,Nehler94,Vader76,Hansteen76,Meyerhof77,Mueller78,Soff79,Graue82,Hansteen85,Rumrich89,Hansteen90,McConnell12,Novak18}
or high-intensity laser pulses%
~\cite{Keldysh65,Faisal73,Reiss92,Corkum93,Gaier02,Salamin06,Milosevic06,Selsto09,DiPiazza12,Pindzola12,Forre14,Kjellsson17}
has been studied for many decades. 
Much less attention has been paid, however, to the ionization in the \textit{combined} field of a projectile and of a laser. 
Here we address the question of whether the simultaneous perturbation of an ion by a charged projectile and a strong laser may lead to remarkable interference effects that modify the ionization probability. 

In order to calculate the probability of the laser-assisted ionization in ion collisions and, hence, to investigate the ``laser + collision'' interference  we have developed a non-perturbative approach, based on solutions of the time-dependent  Dirac equation. 
This equation describes the interaction of an electron with both a laser pulse and Coulomb field of a charged projectile. 
As discussed in Sections~\ref{sub:coulomb}--\ref{sub:laser}, the Coulomb potential of a projectile is taken into account within the monopole approximation, while the coupling to a laser field is treated within the dipole approximation. 
The numerical solution of the Dirac equation and the evaluation of the ionization probability is reviewed then in Section~\ref{sub:combined}. 
The ionization probability, as obtained by using the developed non-perturbative approach, accounts for both Coulomb and laser interactions, as well as to their interference. 
This interference effect is of non-perturbative nature, as demonstrated in Section~\ref{sec:perturbation}. 
Namely, in this section we consider the case of weak laser and projectile potentials, which lead to the perturbative limit where the ionization probability is just an incoherent sum of ``laser only'' and ``collision only'' probabilities. 
Based on this finding we introduce the relative difference between non-perturbative and perturbative probabilities, that is used to quantify the ``laser + Coulomb'' interference effect. 
While the developed theory can be used for a wide range of collision systems, in the present work we restrict our analysis to the laser-assisted scattering of $\alpha$-particles off hydrogen-like lead Pb$^{81+}$ ions. 
The physical parameters and computational details for this system are given in Section~\ref{sec:details}, and the results of the calculations are presented in Section~\ref{sec:results}. 
In particular, we found that the ``laser + Coulomb'' interference may result in significant --- about 10\% --- modification of the ionization probability and, moreover, can be both constructive and destructive depending on the frequency of applied laser pulse. 
Finally, a short summary of these results is given in Section~\ref{sec:conclusions}. 

The relativistic unit system is used, $\hbar = m = c = 1$, unless stated otherwise.

\section{Theoretical background}
\label{sec:theory}
In order to analyze the laser-induced ionization of hydrogen-like ions colliding with bare projectile we have to discuss first the ``building blocks'' of this process. 
Namely, in Sections~\ref{sub:coulomb} and \ref{sub:laser} below we will recall ion-impact-ionization and characterization of the laser pulse, respectively. 
While the theory that accounts for both, collision- and photo-ionization, will be presented in Section~\ref{sub:combined}. 

\subsection{Coulomb ionization in ion-ion collisions}
\label{sub:coulomb}

\begin{figure}
  \includegraphics[width=\columnwidth]{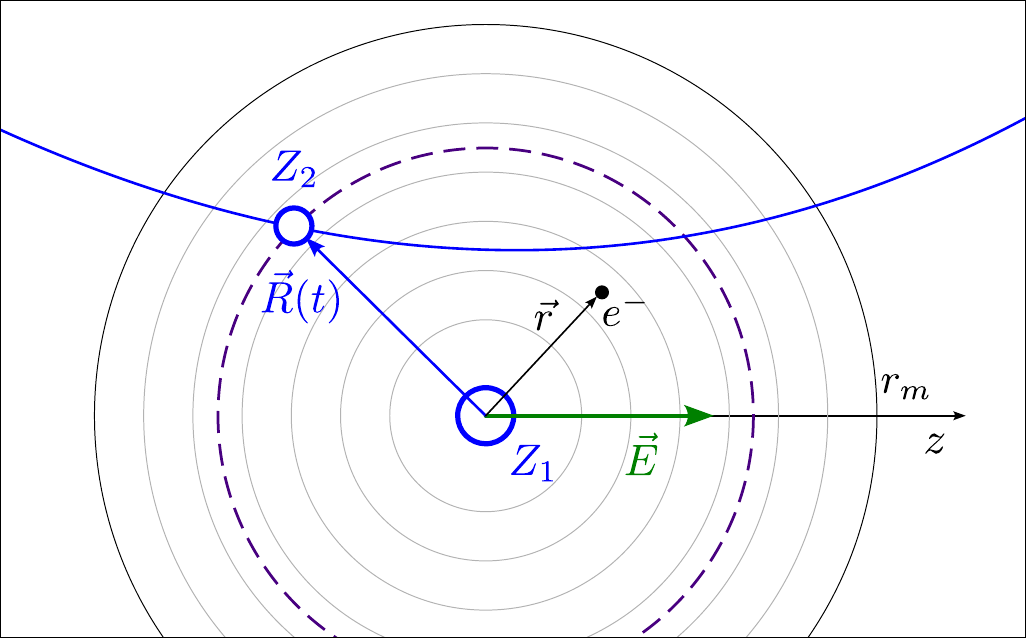}
  \caption{The system layout in the spherical coordinates. $\vec{E}$ is the laser field vector, the dashed circle represent the charged shell of monopole part of the projectile potential, $r_m$ is the boundary for B-spline calculations. The concentric circles illustrate B-spline node boundaries.}
  \label{fig:geometry}
\end{figure}

Below we will briefly remind the basic theory to describe the ionization of a hydrogen-like ion, colliding with a bare nucleus. 
We will describe this process in spherical coordinates, whose origin is located at the nucleus of the target (hydrogen-like) ion, see Fig.~\ref{fig:geometry}. 
The projectile is assumed to be light enough to neglect the recoil of the target nucleus during the collision. 
Moreover, it follows the Rutherford scattering trajectory $\vec R(t)$.
If the collision energy is sufficiently low, one can also neglect the magnetic field of the moving nucleus.

In this study we will describe the Coulomb interaction between the target electron and the projectile ion within the monopole approximation. 
This approximation is known to reproduce the accurate results at small internuclear distances up to $\sim 500$~fm and is widely used to study the processes involving heavy ions \cite{Wille86, Nehler94, Mueller78, Graue82, Rumrich89}.
Ionization takes place predominantly at very small internuclear distances, which makes the monopole basis suitable for numerical calculations.
Within the monopole approach, the electron-projectile interaction is approximated by a potential of a hollow charged shell of radius $R(t)$ centered at the origin of coordinate system.
In this case, the Hamiltonian of an electron in the field of colliding nuclei is given by
\begin{equation}
\label{V01}
  \hat{\Hc}_{0} = \vec\alpha \vec p  + \beta m -\frac{\alpha Z_T}{\max(r, R_T)} - \frac{\alpha Z_P}{\max(r, R(t))},
\end{equation}
where $\alpha$ is the fine structure constant, $\vec \alpha$ and $\beta$ are the Dirac matrices, and $R_T$ is the radius of a target nucleus.
Since the ``monopole'' Hamiltonian Eq.~(\ref{V01}) is spherically symmetric, its eigenfunctions can be conveniently written as 
\begin{equation}
\label{Phi}
	\Phi^{\mu}_{n\varkappa}(\vec r) = \frac{1}{r}
	\begin{pmatrix}
		G_{n\varkappa}(r) \chi^{\mu}_{\varkappa}\\
		iF_{n\varkappa}(r) \chi^{\mu}_{-\varkappa}\\
	\end{pmatrix},
\end{equation}
where, $G_{n\varkappa}(r)$ and $F_{n\varkappa}(r)$ are the real radial functions and
$\chi^{\mu}_{\varkappa}$ denote standard Dirac spin-angular functions.
As usual for the case of spherically-symmetric potentials, electronic states are described by the principal quantum number $n$, the projection $\mu$ of the total angular momentum to the $z$-axis,  and by the spin-orbit quantum number defined as
\begin{equation}
\label{varkappa}
  \varkappa = (-1)^{j+l+\frac{1}{2}} \left(j + \frac{1}{2}\right),
\end{equation}
so that orbital $l$ and total $j$ angular momenta are given by
\begin{equation}
	l = \left\{
	\begin{aligned}
		\varkappa,			&	\: \varkappa > 0, \\
		-\varkappa - 1,	&	\: \varkappa < 0, \\
	\end{aligned}
	\right.
	\qquad
	j = |\varkappa| - \frac{1}{2}.
\end{equation}

The large and small radial components $G_{n\varkappa}(r)$ and $F_{n\varkappa}(r)$ satisfy the set of differential equations
\begin{equation}
\label{dual}
  \begin{array}{l}
    \displaystyle
    \frac{dF_{n\varkappa}}{dr} -\frac{\varkappa}{r} F_{n\varkappa} = -(\Ec-V_C(R,r)-1)G_{n\varkappa},
    \\ \\ \displaystyle 
    \frac{dG_{n\varkappa}}{dr} +\frac{\varkappa}{r} G_{n\varkappa} = -(\Ec-V_C(R,r)+1)F_{n\varkappa},
  \end{array}
\end{equation}
where $V_C(R, r)$ is the Coulomb part of the Hamiltonian $\Hc_0$.
In order to construct these functions we use the standard dual-kinetically balanced B-spline approach~\cite{Johnson88, Shabaev04}.
The calculations were performed using 300 B-splines of order 8 within a box of a size $r_m \approx 300$ relativistic units.
To compute the ionization probability we used a set of about 3500 wave functions $\Phi_n$ corresponding to bound and positive quasicontinuum states with $-1.0 < \Ec_n < 2.5$ and $|\varkappa| \leqslant 8$.

\subsection{Shape of the laser pulse}
\label{sub:laser}

To account for  the laser field, we define its potential in the the Coulomb gauge.
Below we will consider the vector potential $\vec A$ of the form
\begin{equation}
\label{Alas}
    \vec A(\eta) = \frac{\vec e_z E}{\omega} g(\eta) \cos(\eta + \phi),
\end{equation}
where $\eta = \omega t - \vec k \vec r$, $\phi$ is a phase value and  
\begin{equation}
\label{geta}
  g(\eta) = 
  \begin{cases}
    \sin^2\left(\frac{\pi\eta}{\omega T_p}\right),  \qquad &  0 < \eta < \omega T_p, \\
    0 & \mbox{otherwise,}
  \end{cases}
\end{equation}
is an envelope function that describes a short pulse with a duration $T_p$ of a few optical cycles. 
The vector potential Eq.~(\ref{Alas}) describes incident light, linearly polarized along the $z$-axis, with the central frequency $\omega$, the amplitude of the electric field strength $E$, and pulse duration $T_p$.

The theoretical analyis of the laser-induced ionization can be significantly simplified if the electron coupling to the laser pulse is treated within the dipole approximation. 
In this case $\eta \approx \omega t$ and the vector potential Eq.~(\ref{Alas}) is given by: 
\begin{equation}
\label{Adip}
    \vec A(t) = \frac{\vec e_z E	}{\omega} \sin^2\left(\frac{\pi t}{T_p} \right) \cos(\omega t + \phi).
\end{equation}

With approximations made for Coulomb and laser potential, the problem has axial symmetry relative to the $z$-axis.
Therefore, the angular momentum projection on this axis is conserved and the quantum number $\mu$ does not changes during ionization.

\subsection{Solution of the time-dependent problem}
\label{sub:combined}

Having briefly discussed the shape of incident laser pulse and the Dirac equation of an electron in the Coulomb field of two nuclei, we are ready to investigate the laser-induced impact ionization. 
To find the probability of this ionization process, we have to solve the time-dependent Dirac equation of an electron in a combined potential of two nuclei and a laser wave:
\begin{equation}
\label{tdde}
	i\frac{\partial\Psi(\vec r, t)}{\partial t} = 
	\left(\hat{\Hc}_{0} - e \vec\alpha \vec A(t)\right) \Psi(\vec r, t).
\end{equation}
The soulution of Eq.~(\ref{tdde}) is constructed in the form of an ansatz
\begin{equation}
\label{Psi}
    \Psi(\vec r, t) = \sum\limits_\nn a_\nn(t) \Phi_\nn\left( \vec r, R(t) \right),
\end{equation}
where conveniently bold-faced indices denote sets of quantum numbers $\nn = n, \varkappa, \mu$.
Functions $\Phi_\nn\left(\vec r, R(t) \right)$ comprise a quasi-stationary basis, defined at each fixed moment of time $t$ as the eigenfunctions of the two-center Hamiltonian $\hat{\Hc}_{0}$ for the corresponding value of the distance $R$,
\begin{eqnarray}
\label{bde}
    \hat{\Hc}_{0} \Phi_\nn(\vec r, R) = \Ec_\nn \Phi_\nn(\vec r, R).
\end{eqnarray}
For the sake of brevity below we will  omit the arguments of the basis wave functions $\Phi_\nn$.

By substituting the expansion Eq.~(\ref{Psi}) into the Dirac equation Eq.~(\ref{tdde}), we obtain the system of coupled differential equations to determine the amplitudes $a_\nn(t)$: 
\begin{equation}
\label{cce}
    i\dot a_\nn(t) = M_{\nn\kk}  a_\kk(t),
\end{equation}
\begin{equation}
\label{mkndt}
    M_{\nn\kk} = \Ec_\kk\delta_{\nn\kk} 
    -i\left< \Phi_\nn \middle| \dot{\Phi}_\kk\right>
    -e A(t)\left<\Phi_\nn\middle|\alpha_3\middle|\Phi_\kk\right>.
\end{equation}
Taking the derivative with respect to time of the equations
\begin{equation}
	\begin{array}{l}
		\left< \Phi_\nn \middle| \Hc_{0}  \middle| \Phi_\kk\right> = \Ec_\nn \delta_{\nn\kk}, \\ \\
		\left< \Phi_\nn \middle| \Phi_\kk\right> = \delta_{\nn\kk},
	\end{array}
\end{equation}
one can rewrite the matrix element $\left< \Phi_\nn \middle| \dot{\Phi}_\kk\right>$ as~\cite{QEDSF}
\begin{equation}
	\left< \Phi_\nn \middle| \dot{\Phi}_\kk\right> = 
	\frac{  \left< \Phi_\nn \middle| \dot{\Hc}_{0}  \middle |\Phi_\kk\right>   }{\Ec_\kk - \Ec_\nn}.
\end{equation}
Thus, the matrix elements $M_{\nn\kk}$ are
\begin{equation}
\label{cceM}
    M_{\nn\kk} = \Ec_\kk\delta_{\nn\kk} - i \frac{  \left< \Phi_\nn \middle| \dot{\Hc}_{0}  \middle |\Phi_\kk\right>   }{\Ec_\kk - \Ec_\nn}
    -e A(t)\left<\Phi_\nn\middle|\alpha_3\middle|\Phi_\kk\right>.
\end{equation}
To further simplify the system of equations Eqs.~(\ref{cce})--(\ref{cceM}), one may note that the time derivative acts on the Hamiltonian $\Hc_0$ only via its dependence on the internuclear distance $R(t)$.
Hence, we can write this derivative as a sum of its radial and angular parts,
\begin{equation}
\label{ddt}
  \frac{\partial}{\partial t} = \dot R \frac{\partial}{\partial R} - i (\vec\Omega \hat{\vec j}),
\end{equation}
where $\vec\Omega$ is the angular velocity of the internuclear axis and $\hat{\vec j}$ is the electron angular momentum operator.
The matrix elements of the second term in Eq.~(\ref{ddt}) are known to vanish at small internuclear distances \cite{Nehler94}. 
The first matrix element on the rght-hand side of Eq.~(\ref{cceM}) can be written as
\begin{equation}
\label{Mhic}
  \frac{ \left< \Phi_\nn \middle| \dot\Hc_0  \middle |\Phi_\kk\right>    }{\Ec_\kk - \Ec_\nn} = 
	\frac{\dot R }{\Ec_\kk - \Ec_\nn}  \left< \Phi_\nn \middle| \frac{dV_C(R)}{dR}  \middle |\Phi_\kk\right>.
\end{equation}
Solution of the system of coupled equations Eq.~(\ref{cce}) requires evaluation of matrix elements of operator $dV_C(R)/dR$ for different internuclear distances $R$.
Having performed these calculations and by computing the matrix element of operator $\alpha_3$ we are ready to find amplitudes $a_\nn(t)$ numerically.

For the numerical solution of the system of equations Eq.~(\ref{cce}) we split the time into small intervals $\Delta t$.
For each time interval the matrix $M$ is approximated by its middle value,  
$M(t) \approx M(t_i + \Delta t/2)$, $t \in [t_{i}, t_{i+1}]$.
When $M$ is approximated by a constant matrix, the solution of the set of equations Eq.~(\ref{cce}) can be found as
\begin{equation}
  \label{aevo}
  \vec{a}_j(t_i+\Delta t) = e^{-iM\Delta t} \vec{a}_j(t_i).
\end{equation}
However, computation of the matrix exponent may be very demanding. 
Instead, we use the highly efficient Lanczos propagation method to find the vector $\vec{a}(t)$ \cite{Park86}.

After performing the time propagation of the amplitudes $a_\nn$ to $t = \infty$, we can find for the ionization probability
\begin{equation}
\label{wag_def}
  w_{\alpha\gamma} = \sum\limits_{\nn} |a_\nn(t = \infty)|^2,
\end{equation}
where the summation runs over all electronic states, belonging to the positive-energy quasi-continuum, i.e. when $\Ec_\nn>1$.

\section{Perturbative case}
\label{sec:perturbation}

Before discussing the results of the non-perturbative treatment of the laser-induced collisional ionization, it is useful to briefly consider the predictions of the perturbation theory. 
In the case of weak laser and projectile potentials, the amplitudes near the matrix elements in Eq.~(\ref{cceM}) can be approximated as $a_\kk \approx 1$ for the initial state and $a_\kk \rightarrow 0$ otherwise. 
Then, the set of ordinary differential equations Eq.~(\ref{cce}) can be decoupled~\cite{Nehler94, QEDSF}, and we obtain the ordinary differential equation
\begin{equation}
  \label{ccepert}
  i\dot{a}_{\nn} = \Ec_\nn a_{\nn} 
  - i\dot{R} \frac{\langle \Phi_\nn | \frac{dV_C}{dR} | \Phi_\kk \rangle }{\Ec_\nn - \Ec_\kk}  
  -eA(t)\langle \Phi_\nn | \alpha_3 | \Phi_\kk \rangle,
\end{equation}
for the probability amplitude of the transition from the initial state $\kk$ to the final state $\nn$.
In order to analyze the symmetry properties of Eq.~(\ref{ccepert}) we remind that eigensolutions Eq.~(\ref{Phi}) of the spherically-symmetric (monopole) Hamiltonian ${\Hc}_0$ are characterized by the Dirac's angular momentum quantum number $\varkappa$.
We consider ionization from the $1s$ ground state for which $\varkappa = -1$.
A simple angular algebra analysis shows that the matrix element of the operator $dV_C/dR$ has nonzero values only for transitions without change of $\varkappa$.
At the same time, the matrix element of $\alpha_3$ allows transitions to the states with $\varkappa = -2$ and $\varkappa = +1$.
Based on these observations as well as on Eq.~(\ref{wag_def}) one sees that the total ionization probability is given by the sum of ``collision-only'' and ''laser-only'' probabilities, with no interference between these two channels.  
As we will see below, this is not the case for the non-perturbative treatment, where the interference between Coulomb- and laser-ionization terms may remarkably modify the ionization probability.
In order to investigate this interference effect we introduce the relative difference:
\begin{equation}
\label{dw_def}
  \delta w_{\alpha\gamma} = \frac{w_{\alpha\gamma} - (w_\alpha + w_{\gamma})}{w_\alpha + w_{\gamma}}
\end{equation}
between probability $w_{\alpha \gamma}$ of the ionization by a combined ``Coulomb + laser'' potential and the sum of ``Coulomb only'' $w_{\alpha}$ and ``laser only'' $w_{\gamma}$ probabilities. 
The two latter are obtained based on Eq.~(\ref{cceM}) with  the third and second terms omitted, respectively.

\section{Details of calculations}
\label{sec:details}
While the developed approach can be applied for various ion collisions, here we consider ionization of a hydrogen-like lead Pb$^{81+}$ ion by a combined potential of a projectile $\alpha$ particle and a short intense laser pulse.
We will investigate the ionization probability Eq.~(\ref{wag_def}) of this process, which depends on a number of physical parameters which are discussed below.
First, accoding to Eq.~(\ref{Adip}), the laser pulse is defined by its duration $T_p$, frequency $\omega$, maximum field strength $E$, and carrier-envelope phase $\phi$.
To quantify field strength and frequency, we introduce characteristic values
\begin{subequations}
	\label{wesc}
  \begin{align}
		\label{wsc}
		\omega' &= 2\Ec_{bind}, \\ \displaystyle
		\label{esc}
		E' &=  \frac{3\alpha Z_T}{\langle r^2 \rangle},
  \end{align}
\end{subequations}
where $\Ec_{bind}$ is binding energy and $\langle r^2 \rangle$ is the mean square of the radial electron coordinate in the ground state of the target Pb$^{81+}$ ion.
We note that Eqs.~(\ref{wesc}) are similar to the non-relativistic $Z$-scaling of the frequency and the electric field strength~\cite{Selsto09, Gaier02}. 
This scaling, however, is still practical to quantify the strength $E$ and to define weak- and strong-field regimes.

In the present work we perform calculations of the ionization probability for the laser frequency and field strength in the ranges of $0.4 \omega' \leqslant \omega \leqslant 2.0 \omega'$ and $10^{-3} E' \leqslant E \leqslant 1.0 E'$.
For these parameters, the electron-laser coupling can be treated within the dipole approximation. 
To justify the dipole approach we follow Ref.~\cite{Reiss92} and recall that an electron in a field of an electromagnetic wave oscillates along the eight-shaped trajectory. 
The correspoinding displacement, or the amplitude of electron oscillations, can be written as 
\begin{equation}
  X = \frac{1}{2\omega} \frac{U_p}{m+2U_p},
\end{equation}
where $U_p$ is the ponderomotore potential,
\begin{equation}
  U_p = \frac{e^2 E^2}{4m\omega^2}.
\end{equation}
In order to apply the dipole approximation, the amplitude $X$ should be small comparing to the laser wavelenth $\lambda$ as well as to the size of the electron orbit $r_B$.
For the laser frequencies, studied in the present work, the corresponding relations are:
\begin{equation}
  \begin{array}{l}
    2.5\cdot 10^{-9} \leqslant X/\lambda \leqslant 6.3 \cdot 10^{-8}, \\
    1.3\cdot 10^{-8} \leqslant X/r_B \leqslant 1.6\cdot 10^{-6}
  \end{array}
\end{equation}
for the weak field, $E = 0.001 E'$, and 
\begin{equation}
  \begin{array}{l}
    2.4\cdot 10^{-3} \leqslant X/\lambda \leqslant 2.4\cdot 10^{-2}, \\
    1.2\cdot 10^{-2} \leqslant X/r_B \leqslant 0.64
  \end{array}
\end{equation}
for the stronger field regime $E=E'$.
We may conclude, therefore, that the dipole approximation is applicable for our studies, at least when the laser field is not too strong. 
These conclusions are further supported by the comparison between dipole and higher multipole calculations reported in Ref.~\cite{Selsto09}.

Beside the frequency $\omega$, the field strength $E$ and the envelope parameters, one should also agree at which moment of the collision the laser pulse comes. 
To specify this time we introduce a time interval between the moment of the closest nuclei approach $t(R_{min})$ and the moment of the maximum pulse intensity $t(I_{max})$,
\begin{equation} 
\label{tau}
  \tau = t(R_{min}) - t(I_{max}).
\end{equation}  
In the Section~\ref{sec:results} we will investigate how the ionization probability depends on $\tau$. 
The calculations performed for a well-defined time interval (see Eq.~(\ref{tau})) are, however, of theoretical academic interest. 
To investigate a more realistic scenario, we average the ionization probability over some measurement window with respect to the interval $\tau$.
Apparently, the size of the window can not be unambigously defined within the theoretical framework. 
In the present work, we define it as
\begin{equation}
\label{tauint}
  -2T_p \leqslant \tau \leqslant 2T_p,
\end{equation}
where $T_p$ is the pulse duration. 
In the calculations below we set $T_p = 3T$, where $T$ is the period of the optical cycle. 
Moreover, we assume that the carrier envelope phase is zero, $\phi = 0$.

Apart of the laser parameters, discussed above, one also needs to define the impact parameter $\rho$ and the energy $\Ec_{CM}$ that characterize ion-ion collision. 
We performed calculations for collisions with center of mass energy $\Ec_{CM} = 5$ and 10 MeV and impact parameters up to 500~fm.

To conclude the discussion of the numerical procedure, we describe the used technique which allows significant reduction in the computational resources.
Note that on the right hand side of Eq.~(\ref{Mhic}) only $\dot R(t)$ depends on the collision parameters $\Ec_{CM}$ and $\rho$, while the matrix element depends only on the value of $R$ and the type of nuclei.
Similarly, the matrix $\left<\Phi_\nn\middle|\alpha_3\middle|\Phi_\kk\right>$ in Eq.~(\ref{cceM}) does not depend on the laser pulse potential.
This allows us to reuse the calculated matrices for collisions with different collision energies and laser pulse parameters.

\section{Results and discussion}
\label{sec:results}

\begin{figure}
  \includegraphics[width=\columnwidth]{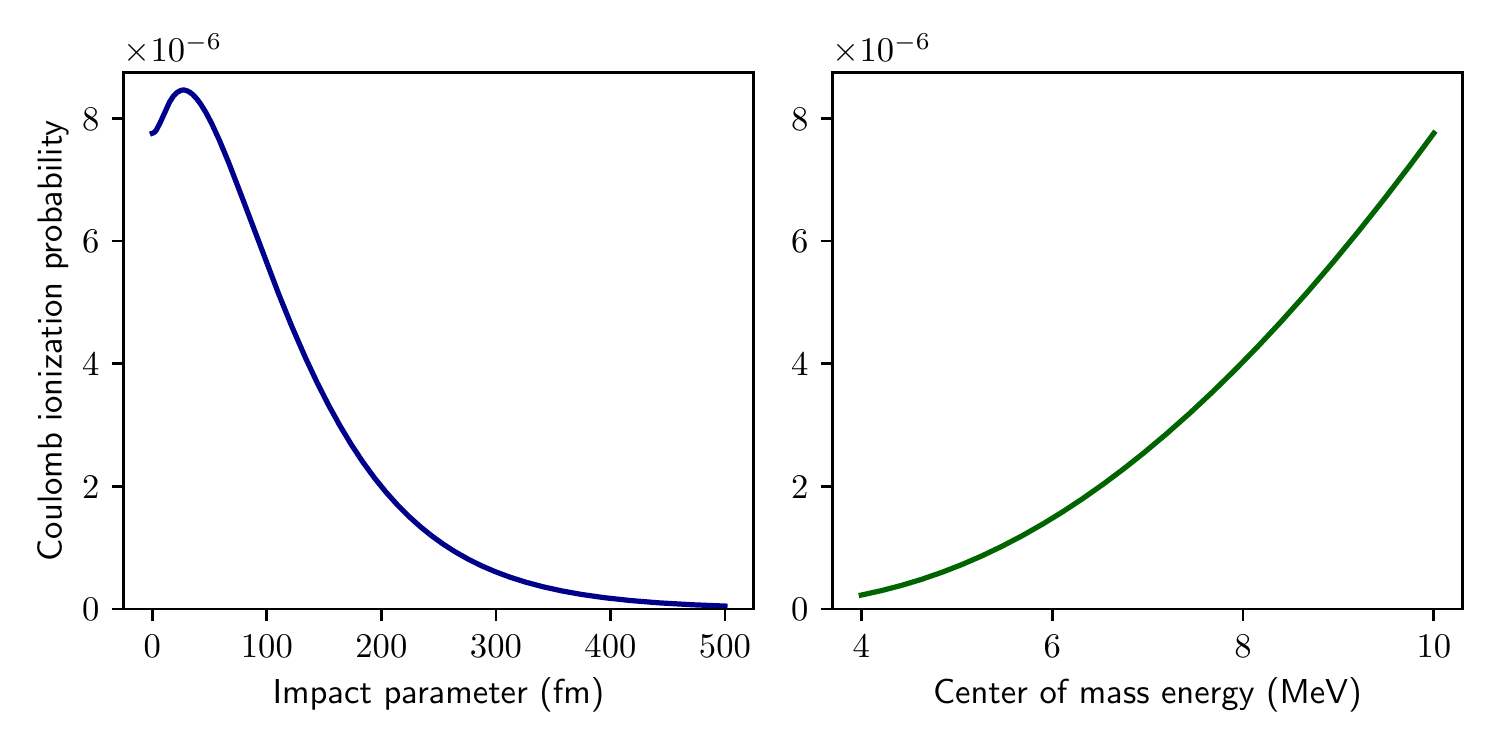}
  \caption{Probability $w_{\alpha}$ of the ground-state ionization of hydrogen-like lead by an $\alpha$ particle impact. 
  The left panel displays  $w_{\alpha}$, calculated for the center of mass energy $\Ec_{CM} = 10$~MeV, as a function of impact parameter $\rho$.
  In contrast, the energy dependence of the ionization probability for the head-on collision, $\rho = 0$, is displayed in the right panel.}
  \label{fig:wa-erdep}
\end{figure}

Before we start our analysis of the ground-state ionization for Pb$^{81+}$ by the combined ``laser + Coulomb'' potential, let us briefly discuss the individual, ``collision only'' $w_{\alpha}$ and ``laser only'' $w_\gamma$ probabilities. 
Figure~\ref{fig:wa-erdep} shows, for example, the probability of the $1s$-ionization by $\alpha$-particle impact. 
The calculations were performed for two scenarious: in the left panel we display $w_\alpha$ as a function of impact parameter $\rho$ but for the fixed center of mass energy $\Ec_{CM} = 10$~MeV, while the energy dependence of $w_{\alpha}$ for the case $\rho = 0$ is presented in the right panel.  
The impact parameter dependence features a local maximum at approximately 50~fm and decreases polynomialy for the higher impact parameters.

We note that the $\rho$-behaviour of $w_{\alpha}$ as well as the monotonic increase of the ionization probability with the center of mass energy, displayed in Fig.~\ref{fig:wa-erdep}, are expected from previous studies \cite{Vader76, Hansteen76, Hansteen85, Hansteen90}.

\begin{figure}
  \includegraphics[width=\columnwidth]{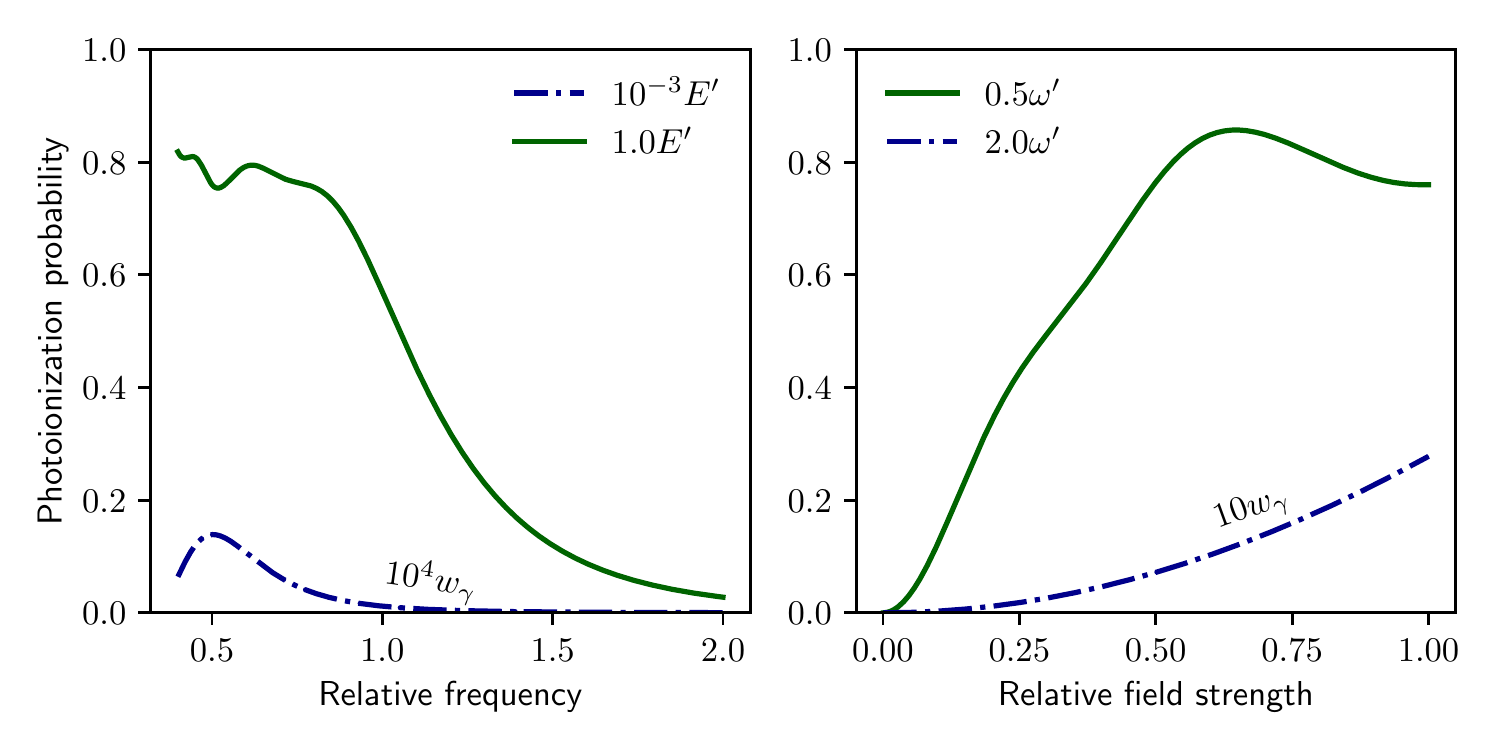}
  \caption{Probability of photoionization of a hydrogen-like Pb by a laser pulse as a function of relative laser frequency $\omega/\omega'$ (left panel) and relative field strength $E/E'$ (right panel). 
  The blue dot-dashed curve is scaled by factor $10^{4}$ in the left panel and by factor $10^1$ in the right panel.}
  \label{fig:wg-wdep}
\end{figure}

In order to investigate the ionization of 1s electron for hydrogen-like lead by a laser pulse, we display in Fig.~\ref{fig:wg-wdep} the ionization probability $w_\gamma$ as a function of (relative) frequency $\omega$ (left panel) and field strength (right panel). 
The calculations of $w_\gamma(\omega)$ were performed for two field strengths, $E = 10^{-3} E'$ (blue dash-dotted line) and $E = 1.0 E'$ (green solid line). 
For the weak field, $w_\gamma$ has maximum near the value of $\omega \approx \Ec_{bind}$ and for both strengths  $w_\gamma$ has an exponential tail at higher frequencies.
In the right panel of Fig.~\ref{fig:wg-wdep} we display $w_\gamma(E)$ as a function of electric field strength and calculated for two laser frequencies, $\omega = 0.5 \omega'$ (green solid line) and $\omega = 2.0 \omega'$ (blue dash dotted line). 
The photoionization probability scales as square of the laser field strength for small $E$~\cite{Selsto09, Pindzola12, Forre14, Kjellsson17}.

\begin{figure}
  \includegraphics[width=\columnwidth]{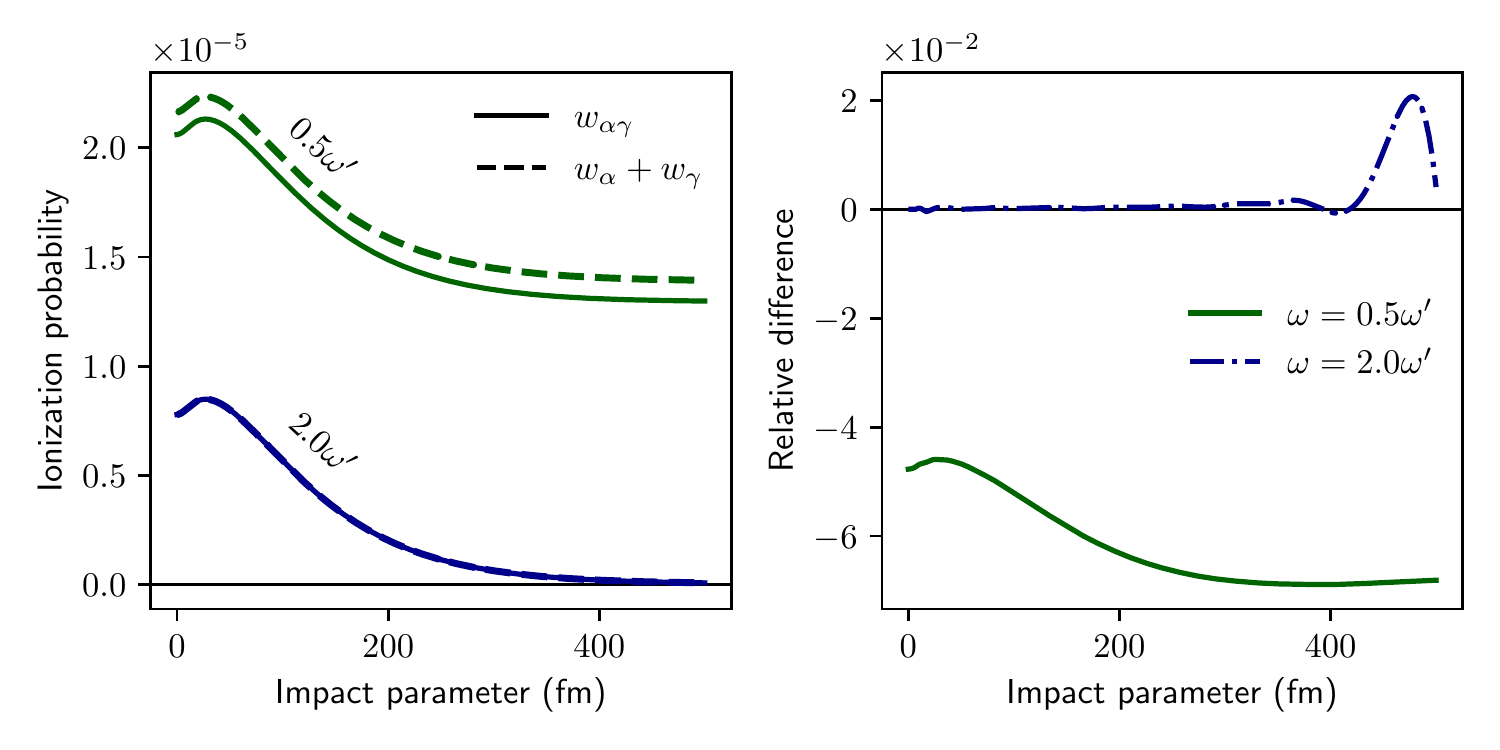}
  \caption{
  Left panel: Probability $w_{\alpha \gamma}$ of the $1s$-ionization of hydrogen-like Pb$^{81+}$ by a combined potential of laser pulse and of $\alpha$ particle as a function of impact parameter. 
  Calculations were performed for the center of mass energy $\Ec_{CM} = 10$~MeV, laser field stregth $E = 10^{-3} E'$ and laser frequencies $\omega = 0.5 \omega'$ (green solid line) and $\omega = 2.0 \omega'$ (blue solid line). 
  Moreover, we compare $w_{\alpha \gamma}$ with the incoherent sum of ``laser only'' and ``collision only'' probabilities (dotted green and blue lines). 
  Right panel: the relative difference $\delta w_{\alpha\gamma}$ between probability of ionization by a combined potential and the sum of ionization probabilities by separate potentials.
  }
  \label{fig:dw-rdep-wk}
\end{figure}

\begin{figure}
  \includegraphics[width=\columnwidth]{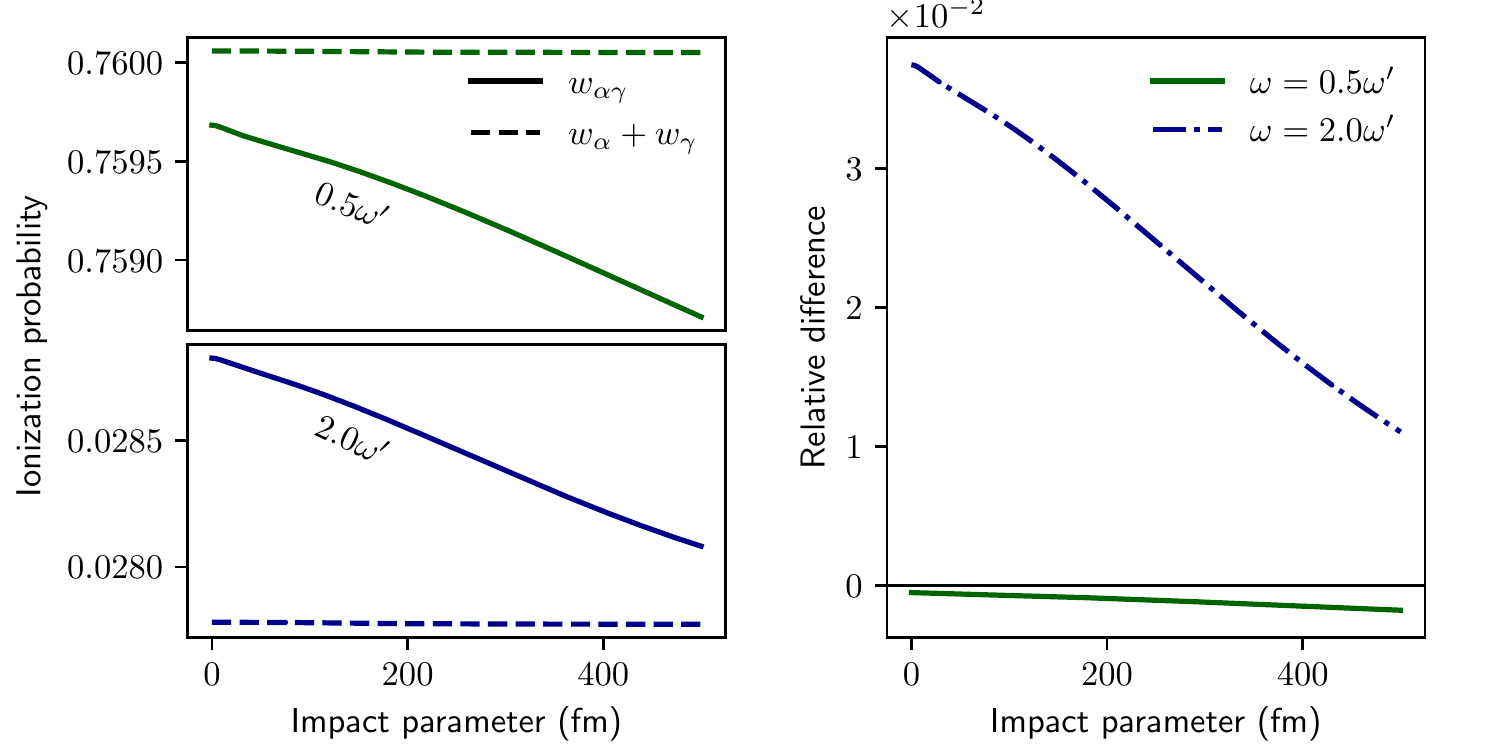}
  \caption{Same as Figure~\ref{fig:dw-rdep-wk} but for the laser field strength $E=E'$.}
  \label{fig:dw-rdep-st}
\end{figure}

\begin{figure}
  \includegraphics[width=\columnwidth]{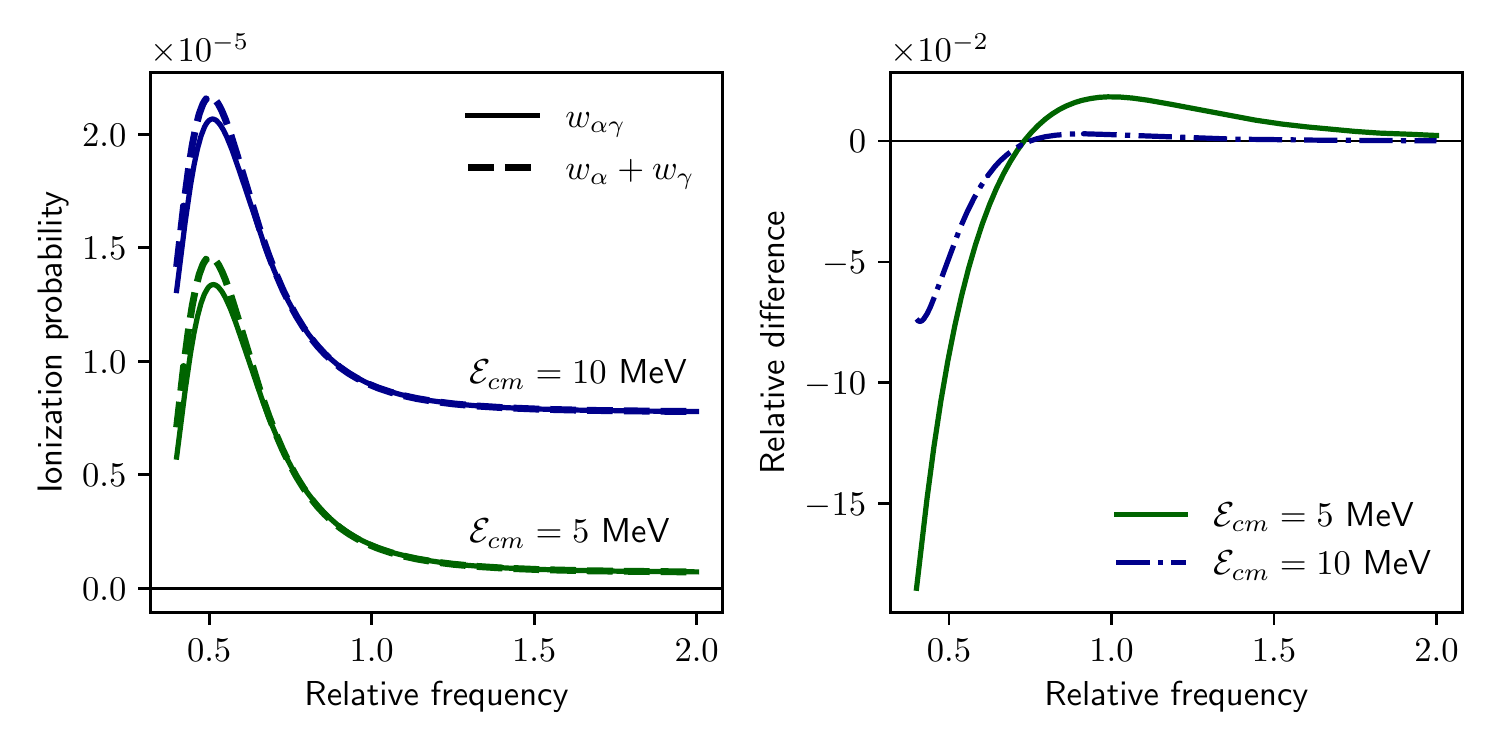}
  \caption{
  Left panel: 
  Probability $w_{\alpha \gamma}$ of the $1s$-ionization of hydrogen-like Pb$^{81+}$ by a combined potential of a laser pulse and of an $\alpha$ particle as a function of relative laser frequency $\omega/\omega'$. Calculations have been performed for the field strength $E=10^{-3}E'$ and the impact parameter $\rho=0$. 
  Moreover, we compare $w_{\alpha \gamma}$ with the incoherent sum of individual ``laser only'' and ``collision only'' probabilities (dashed lines).
  Right panel: The corresponding relative difference Eq.~(\ref{dw_def}).
  }
  \label{fig:w-wdep-wk}
\end{figure}

\begin{figure}
  \includegraphics[width=\columnwidth]{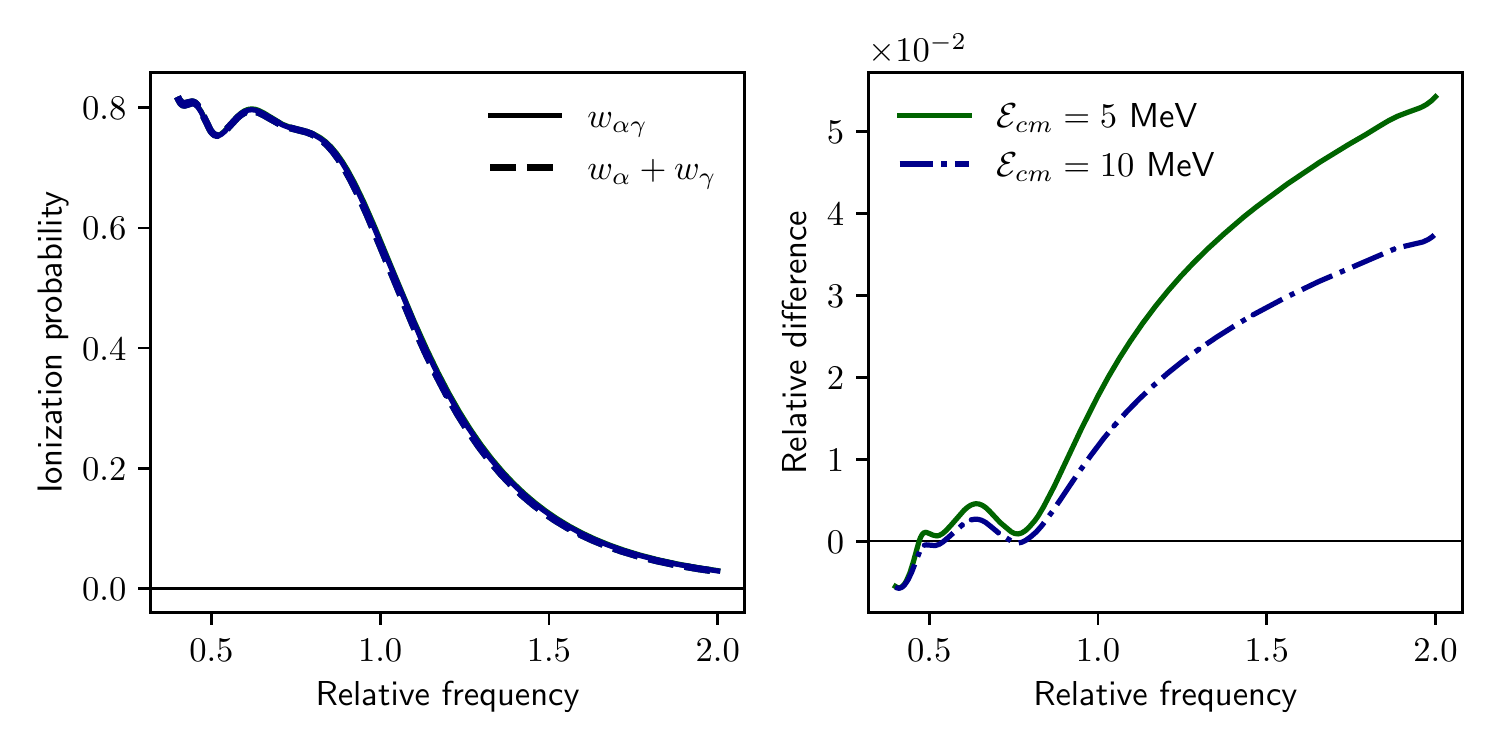}
  \caption{
  Same as Figure~\ref{fig:w-wdep-wk} for the laser field strength $E=E'$.
  }
  \label{fig:w-wdep-st}
\end{figure}

Having briefly discussed the ionization probabilities for the ``laser only'' and ``collision only'' cases, we are ready to explore the ionization by the combined ``laser + Coulomb'' potential. 
The probability $w_{\alpha \gamma}$ is presented as a function of impact parameter $\rho$ in Figs.~\ref{fig:dw-rdep-wk} and \ref{fig:dw-rdep-st}, as well as a function of relative laser frequency $\omega/\omega'$ in Figs.~\ref{fig:w-wdep-wk} and \ref{fig:w-wdep-st}. 
The probability presented in these figures is averaged over the measurement window Eq.~(\ref{tauint}) with respect to the time offset $\tau$, as  explained before.
In order to investigate the effect of the interference between laser and Coulomb ionization channels, we display also the  sum of individual (``laser only'' and ``collision only'') probabilities.
Moreover, the relative difference, as defined in Eq.~(\ref{dw_def}), is displayed in the righ panel of all four figures.  

First, we discuss the $\rho$ dependence of the ionization probability, that is calculated for the center of mass energy $\Ec_{cm} = 10 MeV$, as well as for both laser field strengths, $E = 10^{-3} E'$ (Fig.~\ref{fig:dw-rdep-wk}) and $E = E'$ (Fig.~\ref{fig:dw-rdep-st}), and two laser frequencies, $\omega = 0.5 \omega'$ (green line) and $\omega = 2.0 \omega'$ (blue line). 
As seen from the figures, the interference between laser and Coulomb channels may lead to a remarkable modification of the ionization probability $w_{\alpha \gamma}$ with respect to the incoherent sum $w_{\alpha} + w_{\gamma}$. 
For example, $dw_{\alpha \gamma}$ can reach $\approx 6 \%$ for the low laser frequency and low field strength.

Figs.~\ref{fig:w-wdep-wk}--\ref{fig:w-wdep-st}  allow us to discuss the laser frequency dependence of the ionization probability $w_{\alpha \gamma}$. 
Here, we performed calculations for the head on collisions, $\rho = 0$, for two center of mass energies, $\Ec_{CM} = 5$~MeV and $\Ec_{CM} = 10$~MeV, as well as for two laser field strengthes, $E = 10^{-3} E'$ (Fig.~\ref{fig:w-wdep-wk}) and $E = E'$ (Fig.~\ref{fig:w-wdep-st}). 
Similar to before, we compare here $w_{\alpha \gamma}$ with the sum of ``individual'' probabilities $w_{\alpha}$ + $w_{\gamma}$, and present in the right panels of figures the relative difference Eq.~(\ref{dw_def}). 
As seen from the figures, the ``Coulomb+ laser'' interference may again lead to a remarkable modification of the ionization probability. 
For low laser field strength (Fig.~\ref{fig:w-wdep-wk}) and low frequency $\omega$, for example, the interference results in $\approx 15 \%$ reduction of the $w_{\alpha \gamma}$ when comparing with $w_{\alpha} + w_{\gamma}$. 
In contrast, for high laser field stregth $E= E'$ (Fig.~\ref{fig:w-wdep-st}) and frequency $\omega \gtrsim \omega'$ the interference leads to 3--5 \% enchancement of the ionization probability. 

It could be concluded from the obtained results, that the interference effect has different sign for low and high frequencies.
In the case of strong field, the process is dominated by photoionization, with $w_\gamma \gg w_\alpha$.
Nevertheless, the interaction with the projectile nucleus results in the difference between $w_{\alpha\gamma}$ and the incoherent sum $w_\alpha + w_\gamma$, which is greater than $w_\gamma$ itself.

\begin{figure}
  \includegraphics[width=\columnwidth]{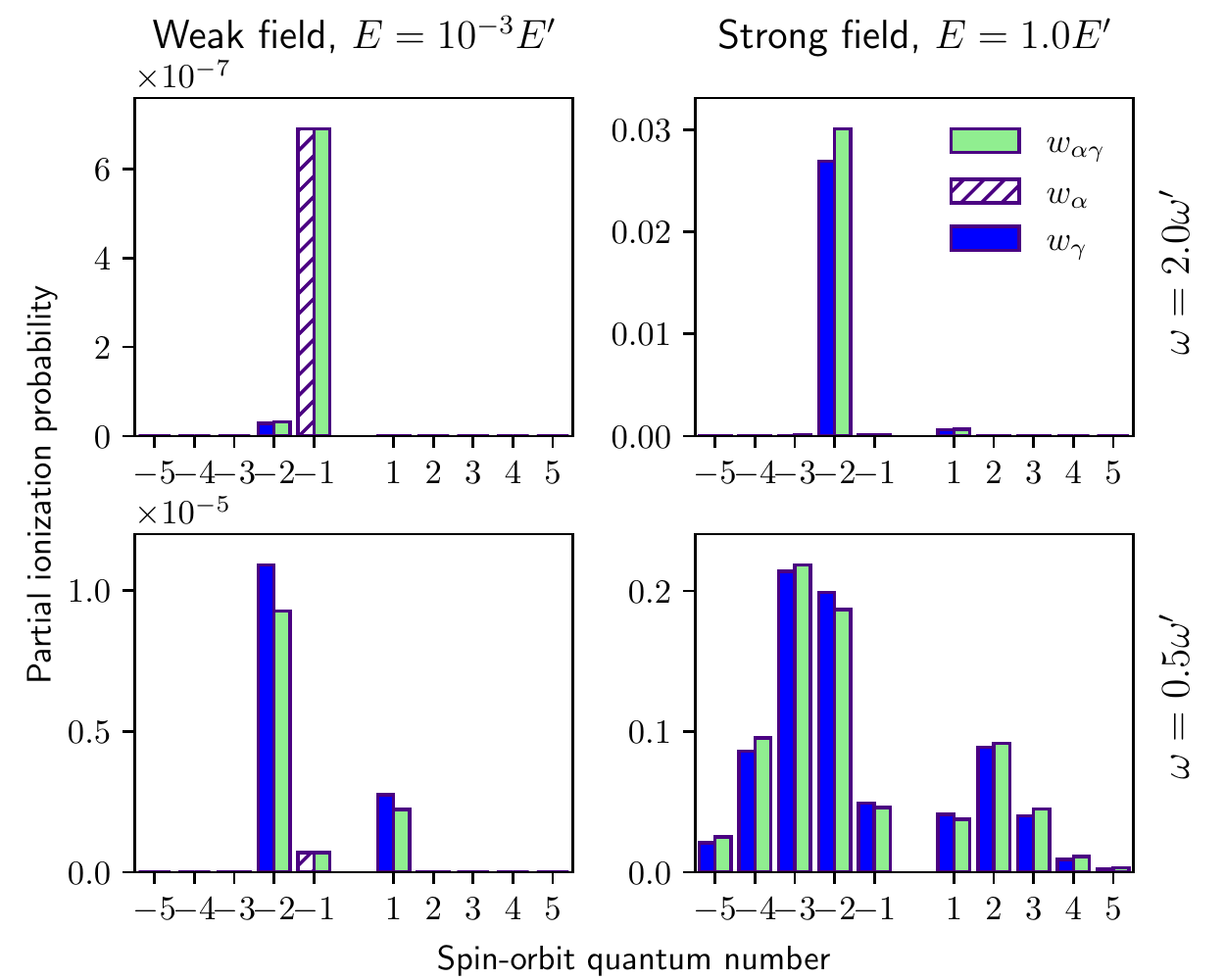}
  \caption{ Partial probability for the ionization of $1s$ electron to the continuum with particular spin-orbit quantum number $\varkappa_f$. 
  The calculations were performed for the center of mass collision energy $\Ec_{CM} = 5$~MeV, laser field strength $E = 10^{-3} E'$ (left panels) and $E = 1.0 E'$ (right panels), as well as for laser frequencies $\omega = 0.5 \omega'$ (lower panels) and $\omega = 2.0 \omega'$ (upper panels). 
  The predictions for the ionization probability $w_{\alpha \gamma}(\varkappa_f)$ for the combined ``laser + Coulomb'' potential (green bar) are compared with those for the incoherent sum of individual channels, $w_{\alpha}(\varkappa_f) + w_{\gamma}(\varkappa_f)$. 
  }
  \label{fig:kspec}
\end{figure}

To better understand the ``laser + Coulomb'' intereference we will consider below the \textit{partial} probabilities:
\begin{equation}
  w(\varkappa_f) = \sum\limits_{n: \: \Ec_n > 1} \left| a_{n \varkappa_f}^\mu \right|^2 \, ,
\end{equation}
for the ionization of an electron into continuum state with a particular angular momentum quantum number $\varkappa_f$.
Figure \ref{fig:kspec} shows the probability of transition of the initially bound $1s$ electron to continuum states with different 
$\varkappa$'s. 
Calculations were done for the particular case of $\tau = 0$, i.e. when the laser pulse intensity reaches its maximum in the moment of the closest nuclear approach.
As seen from the figure, the calculations were done for low- (left column) and high laser field strenth (right column) regimes, as well as for two laser frequencies, $\omega = 2.0 \omega'$ (upper panels) and $\omega = 0.5 \omega'$ (lower panels). 
For all these cases we present the partial ionization probabilities as calculated for the combined ``laser + Coulomb'' potential (right light green bar for each $\varkappa_f$) and for the incoherent sum of ``laser only'' and ``collision only'' channels (left bar, blue color and hatching correspond to contributions of  ``laser only'' and ``collision only'' summands). 
As expected, in the case of weak field and for the ionization from the ground $1s$ state, the main channels are $\varkappa_f = -2,-1,1$, in full compliance with the predictions of the perturbation theory from Sec.~\ref{sec:perturbation}.
In contrast, for the strong-field regime, $E = 1.0 E'$, the population of other continuum states increases, especially for low laser frequency.

Fig.~\ref{fig:kspec} also clearly illustrates the effect of the interference between the photo- and impact ionization channels. 
As seen from the figure, $w_{\alpha \gamma}(\varkappa_f)$ remarkably differs from the sum $w_{\alpha}(\varkappa_f) + w_{\gamma}(\varkappa_f)$ for most of the continuum states $\varkappa_f$. 
For example, for the  weak-field and low-frequency regime (lower left panel), the ``Coulomb--laser'' interference significantly reduces the partial probabilities for $\varkappa_{f} = -2$ and $\varkappa_{f} = +1$ channels. 
In turn, this leads to the reduction of the ``total'' (summed over $\varkappa_f$) probability $w_{\alpha \gamma}$ that was observed in Fig.~\ref{fig:w-wdep-wk}.

\begin{figure}
  \includegraphics[width=\columnwidth]{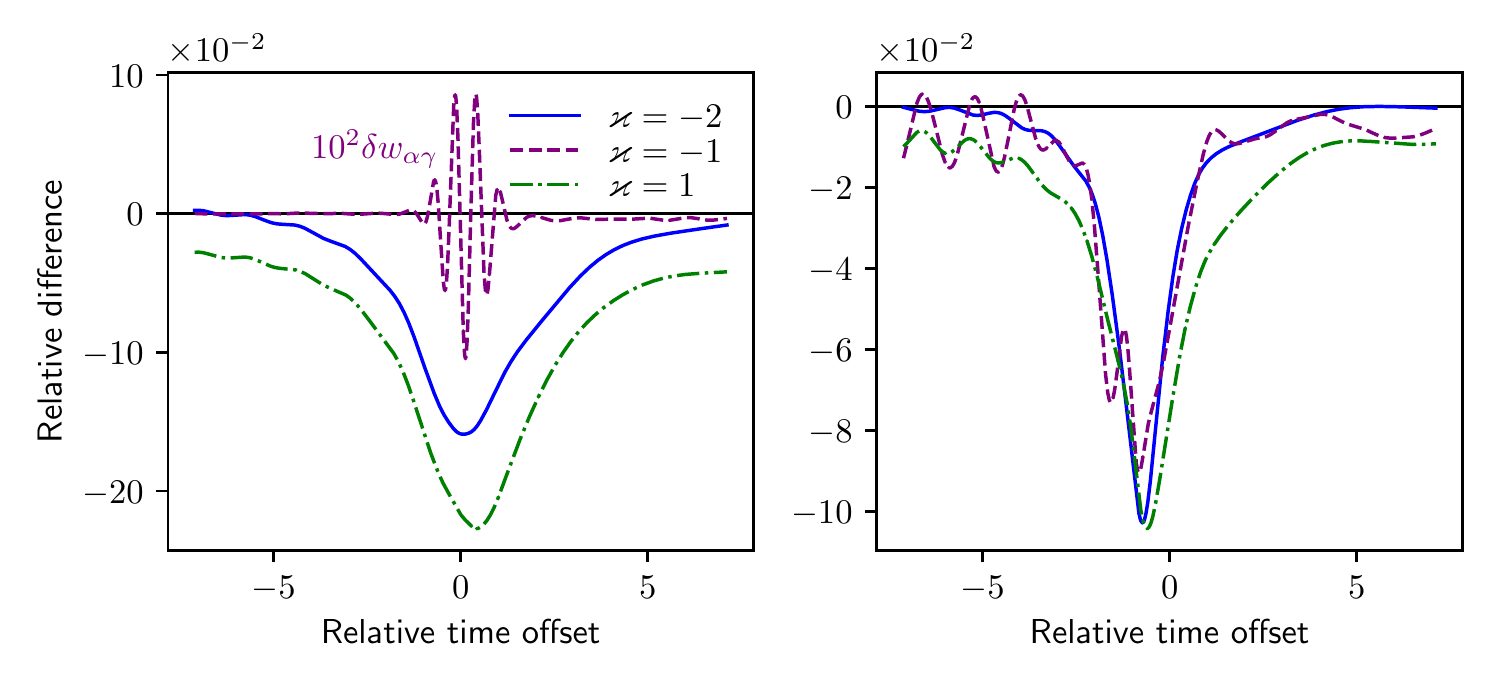}
  \caption{
  The relative difference $\delta w_{\alpha \gamma}(\tau, \varkappa_f)$ for the ionization into particular continuum state $\varkappa_f$ as a function of the time offset $\tau$ between the moments of closest nuclei approach and of laser pulse maximum. The time offset is shown in units of the optical period. 
  The calculations were performed for the collision center of mass energy $\Ec = 10$~MeV, as well as for the laser pulse with the strength $E = 0.001 E'$ (left panel) and $E = 1.0 E'$ (right pannel), and frequency $\omega = 0.5 \omega'$. 
  In the left panel, the curve for $\varkappa_f = -1$ is scaled by factor $10^2$.
  }
\label{fig:dtau.low}
\end{figure}

\begin{figure}
  \includegraphics[width=\columnwidth]{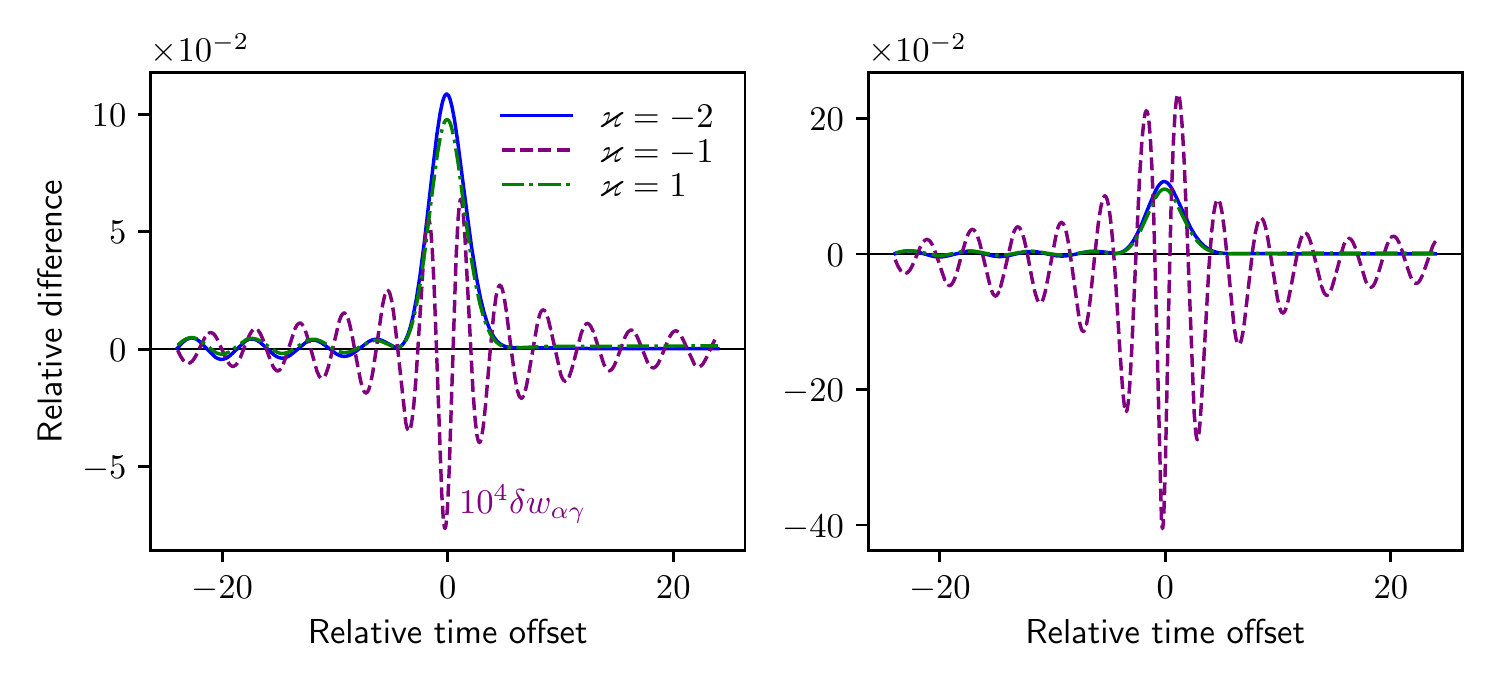}
  \caption{The same as Fig.~\ref{fig:dtau.low} but for the laser frequency $\omega = 2.0\omega'$.
  In the left panel, the curve for $\varkappa_f = -1$ is scaled by factor $10^4$.}
\label{fig:dtau.high}
\end{figure}

Until now, we have investigated the ``Coulomb + laser'' contribution to the ionization probability either for the avearged time offset $\tau$ between the moments of maximal laser pulse intensity and the closest nuclei approach, or for the case $\tau = 0$. 
To better understand how this interference contribution varies with time, in Figs.~\ref{fig:dtau.low} and \ref{fig:dtau.high}  we display the relative difference $\delta w_{\alpha \gamma}(\tau, \varkappa_f)$ as a function of $\tau$ and for particular continuum-electron channels $\varkappa_f = -2, -1, +1$. 
The calculations have been carried out for two laser frequencies, $\omega = 0.5 \omega'$ (Fig.~\ref{fig:dtau.low}) and $\omega = 2.0 \omega'$ (Fig.~\ref{fig:dtau.high}), as well as for the laser field stregthes $E =0.001 E'$ and $E = 1.0 E'$. 

As seen from Fig.~\ref{fig:dw-rdep-st}, the relative difference $\delta w_{\alpha \gamma}(\tau, \varkappa_f)$ and, hence, the ``Coulomb + laser'' interference contribution, is maximal around $\tau = 0$ and tends to zero for large offset times. 
While this qualitative behaviour can be observed for all continuum channels $\varkappa_f = -2, -1, +1$, the quantitative values of  $\delta w_{\alpha \gamma}(\tau, \varkappa_f)$ strongly depend on $\varkappa_f$. 
For example, the relative difference for the continuum states with $\varkappa_f =-2, +1$, which correspond to the photoionization channels in the perturbative limit, can reach about ten percent, and the interference can be either destructive (lower frequency, Fig.~\ref{fig:dtau.low}) or constructive (higher frequency, Fig.~\ref{fig:dtau.high}). 
This behaviour can be observed for both weak and strong field regimes.
In contrast, in the case of the ``collision'' channel with $\varkappa_f = -1$ and weak laser field, the relative difference exhibits oscillatory behaviour and by orders of magnitude smaller than $\delta w_{\alpha \gamma}$'s in the photoionization channels. 
In the case of larger laser strength the relative difference $\delta w_{\alpha \gamma}(\tau, \varkappa_f = -1)$ retains its behaviour but increases its magnitude, as can be seen in Fig.~\ref{fig:dtau.high}(c).

By analyzing the time dependence of the relative difference one can notice small oscillations of $\delta w_{\alpha \gamma}(\tau, \varkappa_f)$ for ``collision'' channels $\varkappa_f = -2, +1$ and for $\tau < 0$; see, for example, left panel of Fig.~\ref{fig:dtau.high}.
To explain this behaviour, we remind that negative time offsets $\tau$ imply that the collision with the projectile happens \emph{before} the arrival of the laser pulse.
In our opinion, the oscillations of $\delta w_{\alpha\gamma}$ can be connected to oscillatory behaviour of the time dependence of ionization probability in ion collisions~\cite{McConnell12}.
Indeed, in the case shown in the left panel of Fig.~\ref{fig:dtau.high}, the pulse duration is much less than the typical collision time.
Therefore, the system can be considered similarly to the pump-probe setup involving two laser pulses~\cite{Drescher02, Krausz09}.

\section{Conclusions}
\label{sec:conclusions}
In conclusion, we presented a theoretical study of electron ionization in laser-assisted ion-ion collisions. 
To calculate the ionization probability, a numerical non-perturbative approach for solution of time-dependent Dirac equation is developed, which accounts for the interaction of the target electron both with a laser pulse and with a Coulomb field of a projectile ion. 
While our non-perturbative approach can be applied to a variety of collision systems, here we considered a particular case of the laser-assisted scattering of $\alpha$ particle by hydrogen-like lead being initially in its ground electronic state. 
Special attention was paid to the question of how the ionization probability can be affected by the \textit{interference} between laser and Coulomb interactions. 
In order to quantify this interference effect we introduced the relative difference between the probability of ionization by a combined (laser + Coulomb) potential and the sum of probabilities of ``laser-only'' and ``collision-only'' processes. 

The calculations of the relative difference were performed for a set of collision and laser parameters, relevant for current experiments. 
Based on these calculations we found that the ``laser + Coulomb'' interference may result in up to 10 \%  modification of the ionization probability and the effect becomes more pronounced for low collision energies $\Ec_{CM}$. 
Moreover, depending on the laser frequency $\omega$ the interference can be both constructive (high $\omega$'s) or destructive (low $\omega$'s). 
These effects can be observed, for example, at the GSI and FAIR facilities in Darmstadt and can provide further insights into laser-induced ion collisions.



\end{document}